\documentclass[]{spie}  

 \usepackage{graphicx} 
\usepackage{amsmath}
\usepackage{epstopdf}
\usepackage[a4paper, total={6in, 8in}]{geometry}
\usepackage{xcolor}
\usepackage{algorithm}
\usepackage{algorithmic}

\usepackage{amsmath,amsfonts,amssymb}
\usepackage{graphicx}
\usepackage[colorlinks=true, allcolors=blue]{hyperref}

\title{Optimization of Petal-Shaped Occulters for Exo-Earth Observation}
\author{Simon Prunet}
\author{Céline Theys}
\author{Claude Aime}
\affil{Université Côte d'Azur, Observatoire de la Côte d'Azur, CNRS, Nice, France}
\date{}
\authorinfo{simon.prunet@univ-cotedazur.fr}

\begin{document} 
\maketitle

\begin{abstract}
Petal-shaped occulters are giant, formation-flying optical devices designed for use in space astronomy. Their primary function is to block the intense starlight to enable the direct observation of exoplanets, a process challenged by the extreme contrast ratio (lower than $10^{-10}$ for an exo-Earth).

This study focuses on optimizing the shape of the occulter in order to significantly reduce Fresnel diffraction. For a circular occulter, the diffraction pattern shows the bright Arago spot. Nevertheless, we  construct an apodized  occulter as the weighted sum of circular occulters of various diameters, owing to the fact  that their diffraction are complex amplitudes which can, with adequate weighting, interfere with each other in such a way as to form a resulting diffraction pattern with a very dark central area.
This basis of circular occulters allows the use of a simple analytical form for their Fresnel diffraction. First, occulter diameters are fixed on a regularly sampled linear scale, and the weights are  obtained by minimizing a quadratic criterion. An improvement is then obtained using a combined optimization over the weights and the diameters sizes.

Eventually, the apodized profile is used to design the petal-shaped occulter for physical implementation  constraints.
The resulting rejection rates are comparable to the results obtained using SISTER (http://sister.caltech.edu), which uses a completely different non-parametric approach.
\end{abstract}


\section{Introduction}

The effectiveness of external-occulting systems is perfectly illustrated by solar eclipses, when the Moon blocks the solar disk and allows the observation of the corona, whose brightness is several million times fainter.

Solar coronagraphs with an external occulter in space already exist. The ESA  Proba-3 experiment with the ASPIICS coronagraph, recently launched, uses two satellites flying in formation 144 meters apart. The occulter spacecraft  carries a $1.42$~m diameter occulting disk, while the coronagraph spacecraft hosts a Lyot coronagraph. It may be seen as a small-scale precursor of Starshades.

Observing exoplanets obviously requires devices at much larger scales. The brightness of an exo-Jupiter or an exo-Earth is  at most $10^{-6}$ to $10^{-10}$ that of its parent star, and the angular separations are well below one arcsecond for the closest targets. A typical Starshade configuration involves a $70$~m diameter occulter placed more than $100{,}000$~km from a $4$~m telescope, to block the starlight and enable the direct detection of exoplanets.

At first glance, these systems might seem straightforward to analyze. However, Fresnel diffraction introduces non-intuitive features. A classic example is the Arago (Poisson) spot produced by a simple circular disk, first emphasized in Poisson's critique of Fresnel's theory more than two centuries ago.
As a result, a purely circular occulter, even at very large scales, is ineffective for exoplanet observations.
Building on Cash's pioneering work \cite{cash06}, the optimal solution is a flower-shaped occulter with carefully designed petals \cite{vand07}.

The present contribution starts from the analytic expression of the complex amplitude of the Arago spot to the optimization of petal-shaped occulters for the detection of exo-Earths. Using a decomposition on a circular occulter basis, we reach a starlight rejection level better than $10^{-12}$. This performance is comparable to results published with the SISTER software suite \cite{hildebrandt2021}, that were obtained via a totally different approach \cite{vand07}, by enforcing  a $10^{-10}$ rejection over the full telescope pupil.

\begin{figure}[h]
    \centering
      \includegraphics[width=0.6\linewidth]{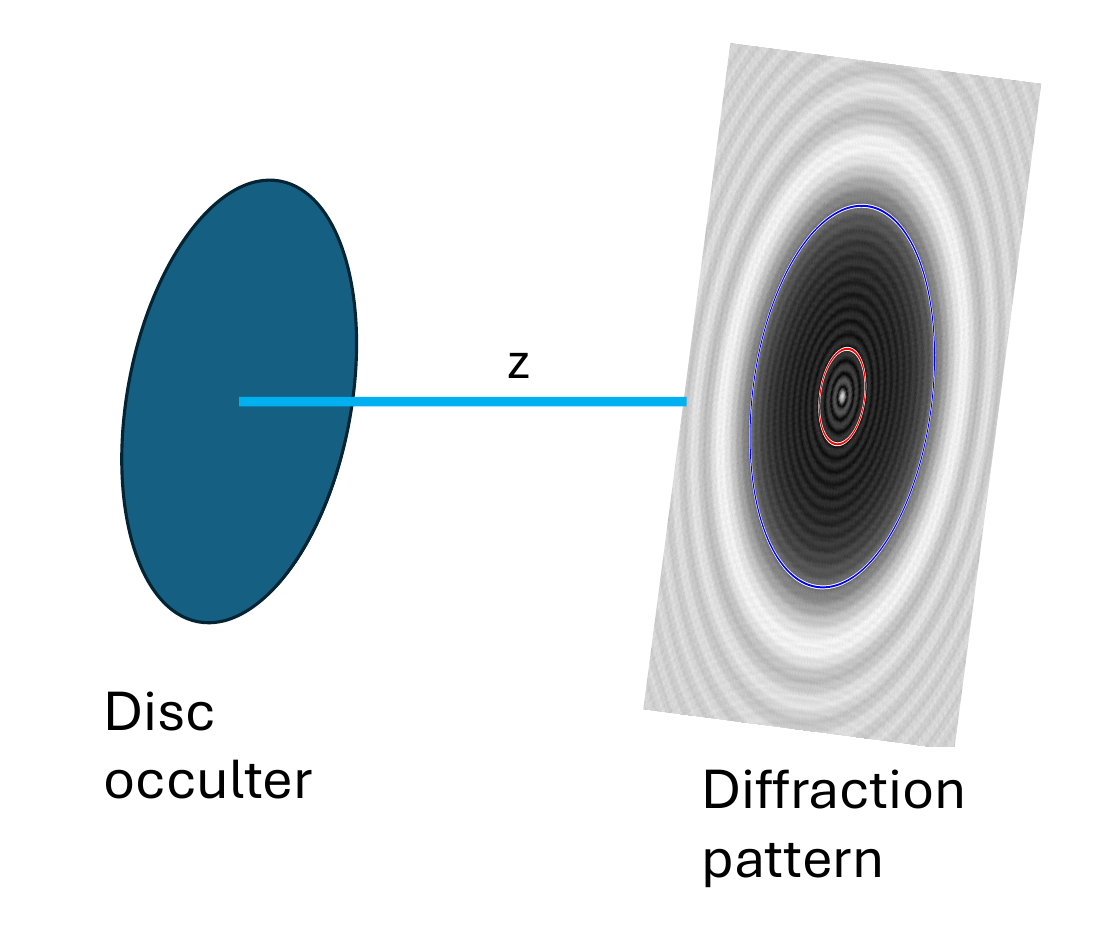}
    \caption{Schematic diagram of an external-occulting coronagraph using a simple disk. The diffraction pattern is shown in the telescope aperture plane; the red circle indicates the pupil, perturbed at its center by the Arago spot.}
    \label{Fig:manip}
\end{figure}
\begin{figure}[h]
    \centering
       \includegraphics[width=0.6\linewidth]{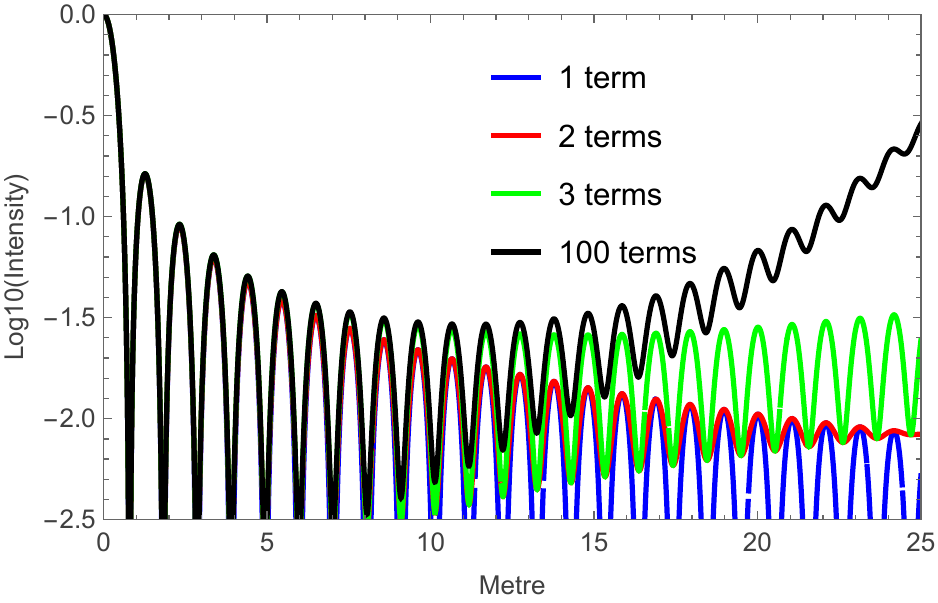}\\
       \includegraphics[width=0.62\linewidth]{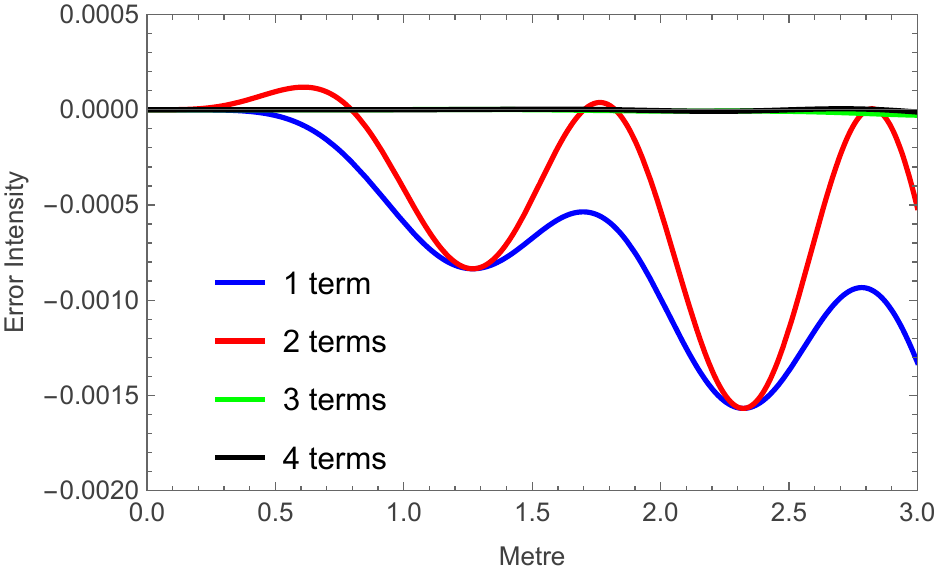}
    \caption{Top panel: cross-section of the intensity produced by an occulting disk, $R=25$~m, $z=80000$~km, $\lambda=650$ nm, computed  using $1$, $2$, $3$ and $100$ terms of the Lommel series (Eq.~\ref{Eq:approx}). Bottom panel: Intensity difference between the computation using $1$, $2$, 3 and 4 terms respectively, and the result obtained with $100$ terms. }
    \label{Fig:Compapprox}
\end{figure}

\section{Fresnel diffraction of an occulter}
Figure~\ref{Fig:manip} provides a schematic diagram of an external-occulting coronagraph using a disk. The bright structure at the center of the shadow is the Arago spot, which is the starting point of our study.

The star is modeled as a point source at infinity producing a plane wave at the occulter. The complex amplitude diffracted by a general occulter with transmission $t(x,y)$ at distance $z$ and wavelength $\lambda$ can be written, within the Fresnel approximation, as the spatial convolution of the occulter by a quadratic phase term normalized by $i\lambda z$:
\begin{equation}
    \begin{split}
        \psi(x,y) &=t(x,y)   * \frac{1}{i \lambda z}\exp(i \pi \dfrac{ x^2+y^2}{\lambda z})\\
        &=1-f(x,y)   * \frac{1}{i \lambda z}\exp(i \pi \dfrac{ x^2+y^2}{\lambda z}),
    \end{split}
    \label{eq:Fresnel}
\end{equation}
where $f(x,y)=1-t(x,y)$ is the complementary screen. This property is similar to Babinet's theorem. The advantage of this formulation is that $f(x,y)$ is bounded by the occulter size. To date, all considered transmissions are real between 0 and 1 (no phase occulter).


For a circular symmetric occulter
of variable transmission
 $f(r)$, where $r=\sqrt{x^2+y^2}$, the complex diffraction amplitude can be written as:
\begin{equation}
    \label{Eq:Base}
    \psi(\rho, \lambda)=1-\frac{\tau_z(\rho) }{i \lambda z}  \int_0^{R}  2 \pi  r f(r) \tau_z(r) J_0(2 \pi \frac{r \rho}{\lambda z}) dr,
\end{equation}
\noindent where $\tau_z(r)= \exp(i \pi r^2 / \lambda z) $ is the quadratic phase term and $J_0$ is the Bessel function of the first kind. The radial coordinate is $r$ in the occulter plane and $\rho$ in the telescope pupil plane.

In the general case, evaluating Eq.~\eqref{Eq:Base} is not analytic. In the particular case of an occulting disk of radius $R$ ($f(r)=1$ for $0\leq r \leq R$ and $0$ elsewhere),  $\psi(\rho,\lambda)$, for $\rho<R$, can be written as  the Lommel series \cite{aime2013theoretical}:
\begin{equation}
    \psi(\rho,\lambda)=\tau_z(R)\tau_z(\rho) \sum _{n=0}^{\infty }
    \left(\frac{\rho}{i R}\right)^n J_n\left(\frac{2 \pi  \rho R}{z \lambda
   }\right)
   \label{Eq:approx}
\end{equation}
In this  presentation, for the sake of clarity,  we use only the first term of the series which  provide a fair approximation of the complex amplitude at the center area of the shadow. Using several terms is nevertheless easy to do and does not change the optimization approach developed here. Note that using 4 terms in the Lommel series will meet all the requirements for a Starshade.

Figure \ref{Fig:Compapprox}, top panel,  compares the result of the series computed for 100 terms with those obtained with just  1, 2  and 3 terms  of Eq.~\ref{Eq:approx}. The comparison is shown for the diffracted intensity (squared modulus of the complex amplitude). The central part is the Arago spot, whose first zero is set by the Bessel function and occurs at $\rho \sim 0.38 \lambda z/R$, i.e. about $80$~cm from the center for the parameters $z$, $\lambda$, and $R$ given in the figure caption. The intensity difference between computations using 1 to 4 terms and the full 100 terms are given in the bottom panel of the figure.
Note that the center of the Arago spot has the same intensity (here 1) as the incident wave.


\section{Apodized occulter}

We partly follow the presentation made in Flamary \emph{et al.}\cite{flamary2014optimization}, where the function $f(r)$ was decomposed on a basis of trapezoidal functions, whose Fresnel diffraction patterns are not analytical. The computations of Eq.~\ref{Eq:Base} were performed numerically there.

As already indicated, in the present study we use a decomposition on a basis of circular occulters, using disc functions, whose Fresnel diffraction is defined in Eq.~\ref{Eq:approx}, and we restrict ourselves to the first term, enabling very fast computations that facilitates the exploration of parameters such as occulter sizes and separations, as well as petal size.

\begin{figure}
    \centering
    \includegraphics[width=0.5\linewidth]{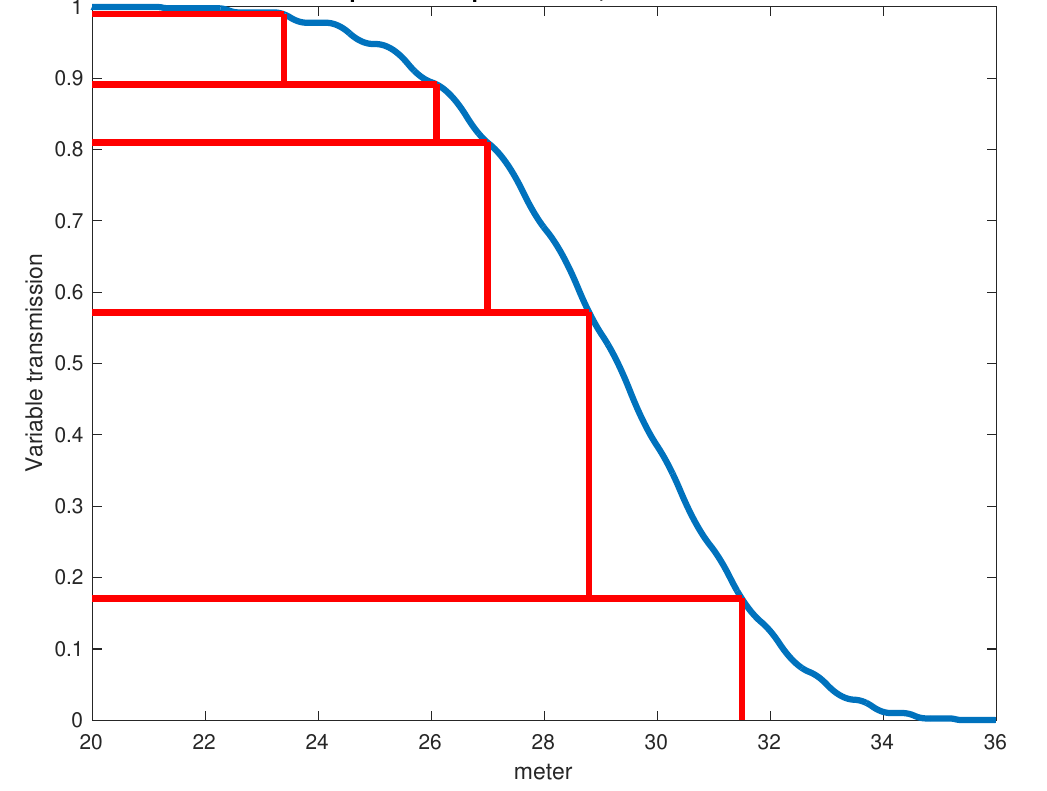}
    \caption{Example of a function $f(r)$ (blue) approximated by $K=5$ disc functions $\Pi\!\left(\frac{r}{R_k}\right)$ (red), see Eq.~\eqref{Eq:somme de Pi}.}
    \label{fig:sommePortes}
\end{figure}

The decomposition of $f(r)$ then becomes:
\begin{equation}
    f(r)=\sum_{k=1}^K \alpha_k f_k(r)=\sum_{k=1}^K \alpha_k \Pi(\frac{r}{R_k})
    \label{Eq:somme de Pi}
\end{equation}
\noindent where the $R_k$ are a set of $K$ radii, as illustrated in the simplified example of Fig.~\ref{fig:sommePortes}.

With this box-function decomposition, the complex amplitude at point $\rho$ can be written as a weighted sum of the functions $\psi_k(\rho,\lambda)$ (first term of Eq.~\ref{Eq:approx}):
 \begin{equation}
  \psi(\rho,\lambda) =\sum_{k=1}^K \alpha_k \psi_k(\rho,\lambda)
  \end{equation}
  i.e.,
\begin{equation}
     \psi(\rho,\lambda) = \exp(i \pi \frac{\rho^2}{\lambda z})\sum_{k=1}^K \alpha_k\exp(i \pi \frac{R_k^2}{\lambda z}) J_0(2 \pi \frac{\rho R_k}{\lambda z}),
         \label{Eq:approxSum}
\end{equation}
and its intensity:
 \begin{equation}
 I(\rho,\lambda)=\big | \psi(\rho,\lambda)\big |^2 =\big | \sum_{k=1}^K \alpha_k \exp(i \pi \frac{R_k^2}{\lambda z}) J_0(2 \pi \frac{\rho R_k}{\lambda z})\big |^2,
\end{equation}
that is,
 \begin{equation}
    I(\rho,\lambda) =\sum_{k,l=1}^{K,K}\alpha_k \alpha_l \cos(\pi \frac{R_k^2-R_l^2}{\lambda z})J_0(2 \pi \frac{\rho R_k}{\lambda z})J_0(2 \pi \frac{\rho R_l}{\lambda z}).
    \label{quad}
    \end{equation}    
Eq.~\ref{quad} is a quadratic form in $\boldsymbol{\alpha}$ that can be written in matrix form as:
     \begin{equation}
     I(\rho,\lambda) =\boldsymbol{\alpha}^T \textbf{K}(\rho,\lambda) \boldsymbol{\alpha}
     \label{I}
\end{equation}
with $\textbf{K}_{k,l}(\rho,\lambda)=\left[\cos\!\left(\pi \frac{R_k^2-R_l^2}{\lambda z}\right)J_0\!\left(2 \pi \frac{\rho R_k}{\lambda z}\right)J_0\!\left(2 \pi \frac{\rho R_l}{\lambda z}\right)\right]$ and $\boldsymbol{\alpha}^T=[\alpha_1 \alpha_2 \ldots \alpha_K]$.


Then the residual flux in a central circular region of radius $\rho_M$ (with $\rho_M$ larger than the telescope radius):

\begin{equation}
 I(\lambda)=\boldsymbol{\alpha}^T \left( \int_{0}^{\rho_M} 2 \pi \rho \textbf{K}(\rho,\lambda) \text{d}\rho \right)\boldsymbol{\alpha}.
 \label{Ilambda}
\end{equation}
In matrix notation:

\begin{equation}
 I(\lambda)=\boldsymbol{\alpha}^T\mathbf{K}_{\rho_M}(\lambda)\boldsymbol{\alpha},
\label{Gamma}
\end{equation}
where $\textbf{K}(\rho,\lambda) $ and $\textbf{K}_{\rho_M}(\lambda)$ are $K\times K$ matrices. 
\section{An optimization problem}

\label{sec:optim}
This part follows the approach of Flamary \emph{et al.}\cite{flamary2014optimization}. We now formulate the design of the apodized occulter as a constrained optimization problem. The apodization profile is parameterized by the coefficient vector $\boldsymbol{\alpha}$ and by the 
radii $\{R_k\}$ that define the piecewise-constant function. The criterion to minimize is the residual stellar flux integrated over the central region of diffraction $I(\lambda)$.
\subsection{Optimization of the weights}
 In a first step, we consider only the optimization of $\boldsymbol{\alpha}$ for fixed radii $\{R_k\}$:
    \begin{equation}
        \min_{\boldsymbol{\alpha}} \; \; \boldsymbol{\alpha}^T(\textbf{K}_{\rho_M} +\mu \mathbf{I})\boldsymbol{\alpha}.
        \label{Min1}
    \end{equation}
Since $\textbf{K}_{\rho_M}$ is ill-conditioned, adding $\mu$ to its diagonal improves numerical stability. This is equivalent to introducing a regularization term $\mu\|\boldsymbol{\alpha}\|^2$ and thus enforcing smoothness on $\boldsymbol{\alpha}$.

Additional conditions on $\boldsymbol{\alpha}$ apply, illustrated in Fig.~\ref{fig:sommePortes}. Since the $K$ box functions are all equal to 1 for $r\leq R_1$, we obtain the condition $\sum_k\alpha_k=1$. Then, for $R_1 \leq r \leq R_2$, the condition becomes $0\leq \sum_{k=2}^{K}\alpha_k \leq 1$, and so on up to $0\leq \alpha_K \leq 1$ for $R_{K-1} \leq r \leq R_K$.

These conditions can be written in matrix form, $\boldsymbol{A}\boldsymbol{\alpha}\geq \boldsymbol{b}$, with:

    \begin{equation}
        \boldsymbol{A}=\left[\begin{array}{ccccc}
             1&1&\ldots&1&1 \\
             0&1 &\ldots&1&1 \\
             \vdots & \vdots& \ddots & \vdots & \vdots \\
             0&\ldots &\ldots & 0 & 1 \\
             -1&-1&\ldots&-1&-1 \\
             0&-1 &\ldots&-1&-1 \\
             \vdots & \vdots& \ddots & \vdots & \vdots \\
             0&\ldots &\ldots & 0 & -1
              \end{array}\right]
              \;\;\;  \boldsymbol{b}=\left[\begin{array}{c}
             0 \\
             0\\
             \vdots\\
             0 \\
             -1 \\
             -1 \\
             \vdots \\
             -1
              \end{array}\right]
    \end{equation}
The problem of Eq.~\ref{Min1} then becomes:
\begin{align}
     &\min_{\boldsymbol{\alpha}} \; \; \boldsymbol{\alpha}^T(\textbf{K}_{\rho_M} +\mu I)\boldsymbol{\alpha} \\
& \text{s.t.} \;\;\boldsymbol{A}\boldsymbol{\alpha}\geq \boldsymbol{b},\;\;\;\mathbf{1}^T\boldsymbol{\alpha}=1
        \label{MinCont}
    \end{align}
For fixed radii $\mathbf{R}$, this is a \emph{convex} quadratic problem because the objective is a quadratic form with a symmetric positive semidefinite matrix (and adding $\mu>0$ makes it strictly convex).
For this quadratic optimization problem with equality and inequality constraints, several methods exist in the literature \cite{boyd2004convex}. The inequality constraint excludes any analytic solution such as the one proposed in Wasylkiwskyj and Shiri\cite{Wasylkiwskyj:11}.

\begin{figure}
    \centering
    \includegraphics[width=0.8\linewidth]{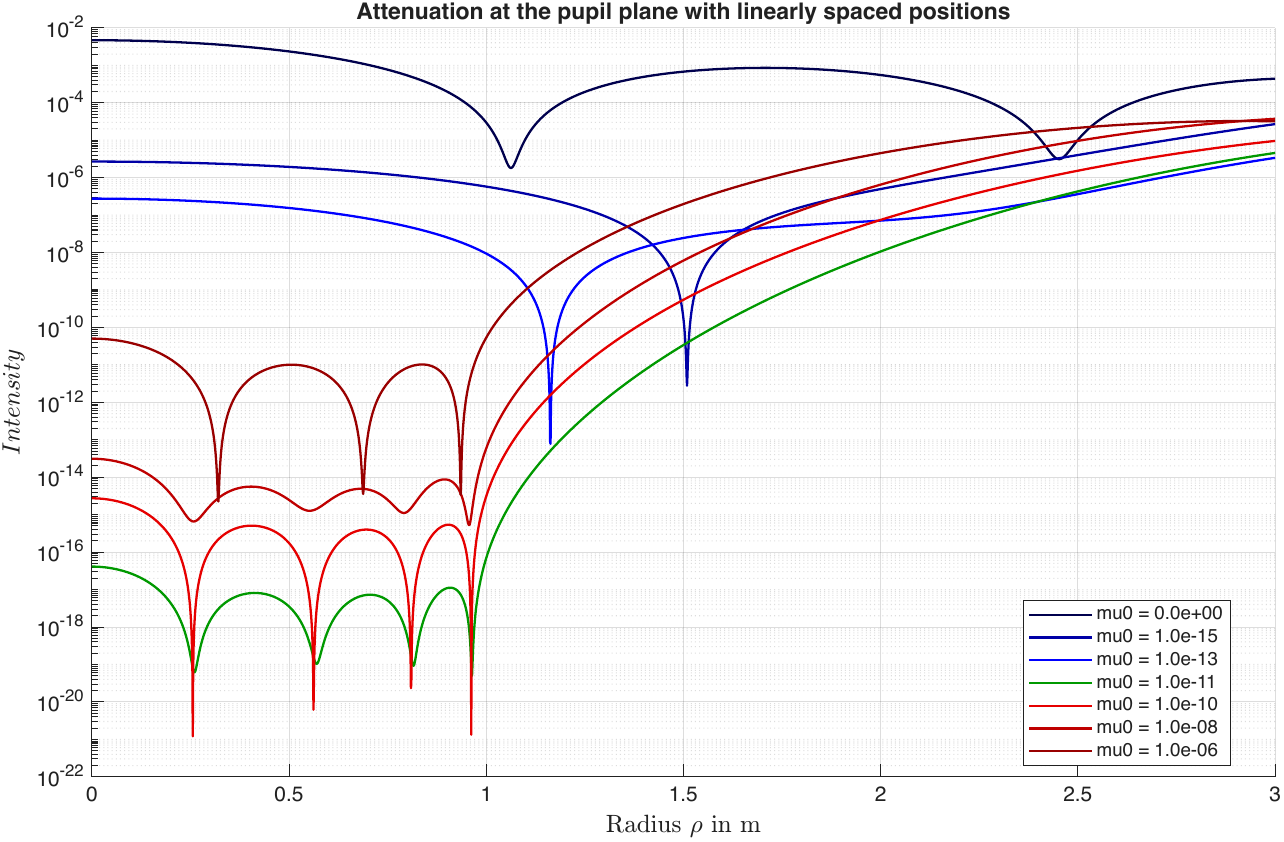}
    \caption{Residual intensity level (umbra)  as a result of the optimization of the $K=1200$ coefficients $\alpha_k$, for various values of $\mu$. The importance of this regularization term is fundamental. The best result is obtained here for $\mu=10^{-11}\max(\textbf{K}_{\rho_M})$.}
    \label{fig:reseau}
\end{figure}

We used MATLAB's \texttt{quadprog}, which implements an interior-point algorithm. To ensure the accuracy required to reach the lowest possible residual flux, the number of iterations and the precision were significantly increased. An example of the residual flux after optimization of $\boldsymbol{\alpha}$ with $K=1200$ box functions is shown in Fig.~\ref{fig:reseau} for various values of $\mu$.

One can see from this set of curves the effect of the parameter $\mu$: the residual intensity decreases as $\mu$ increases, until it reaches a minimum at an optimal value, $\mu=10^{-11}$ here, corresponding to the green curve; beyond this point, the intensity increases again.

\subsection{Optimization of the weights and radii}
To determine the optimal radii $\mathbf{R}$ with $\boldsymbol{\alpha}$ fixed, we  minimize \eqref{Ilambda} with respect to $\mathbf{R}$:
\begin{equation}
\min_{\mathbf{R}} I(\lambda)
\label{min2}
\end{equation}

Although Eq.~\ref{min2} defines a well-posed optimization, the minimization with respect to the radii $\mathbf{R}$ is, in general, \emph{non-convex} because $I(\lambda)$ depends on oscillatory terms (cosines and Bessel functions) of $R_k$. In practice, we nevertheless observed a stable decrease of the criterion when starting from an evenly spaced initialization of the $R$'s (within $[R_{\min},R_{\max}]$) and performing a few gradient-based refinement steps. For that, we compute:
\begin{equation}
\frac{\partial I}{\partial R_m}=\sum_{k,l}\alpha_k \alpha_l \frac{\partial\textbf{K}_{k,l}(\lambda)}{\partial R_m}
\label{grad}
\end{equation}
This calculation yields a closed-form expression, with the derivation steps provided in the Appendix.

Since the gradient can be computed analytically, a gradient-descent algorithm can be used:

\begin{equation}
R_k^{new}=R_k^{old}-\eta \frac{\partial I}{\partial R_k},
\end{equation}
where $\eta$ is a step size to choose such that $I^{new}(\lambda) < I^{old}(\lambda)$. 

We performed this minimization using MATLAB’s \texttt{fmincon}, supplying the analytical gradient.
This leads to the alternating optimization algorithm in Algorithm~\ref{double}.
\begin{algorithm}
\caption{Alternating optimization algorithm}\label{double}
\begin{algorithmic} 
\WHILE{$I^{new}(\lambda) \geq I^{old}(\lambda)$}
\STATE $\boldsymbol{\alpha} \leftarrow \arg \min_{\boldsymbol{\alpha}} I(\lambda)$ with \texttt{quadprog} 
\STATE $\mathbf{R_k} \leftarrow \arg \min_{\mathbf{R_k}} I(\lambda)$ with \texttt{fmincon} 
\STATE $I^{new}(\lambda) \leftarrow I^{old}(\lambda)$
\ENDWHILE
\end{algorithmic}
\end{algorithm}

\begin{figure}
    \centering
    \includegraphics[width=0.7\linewidth]{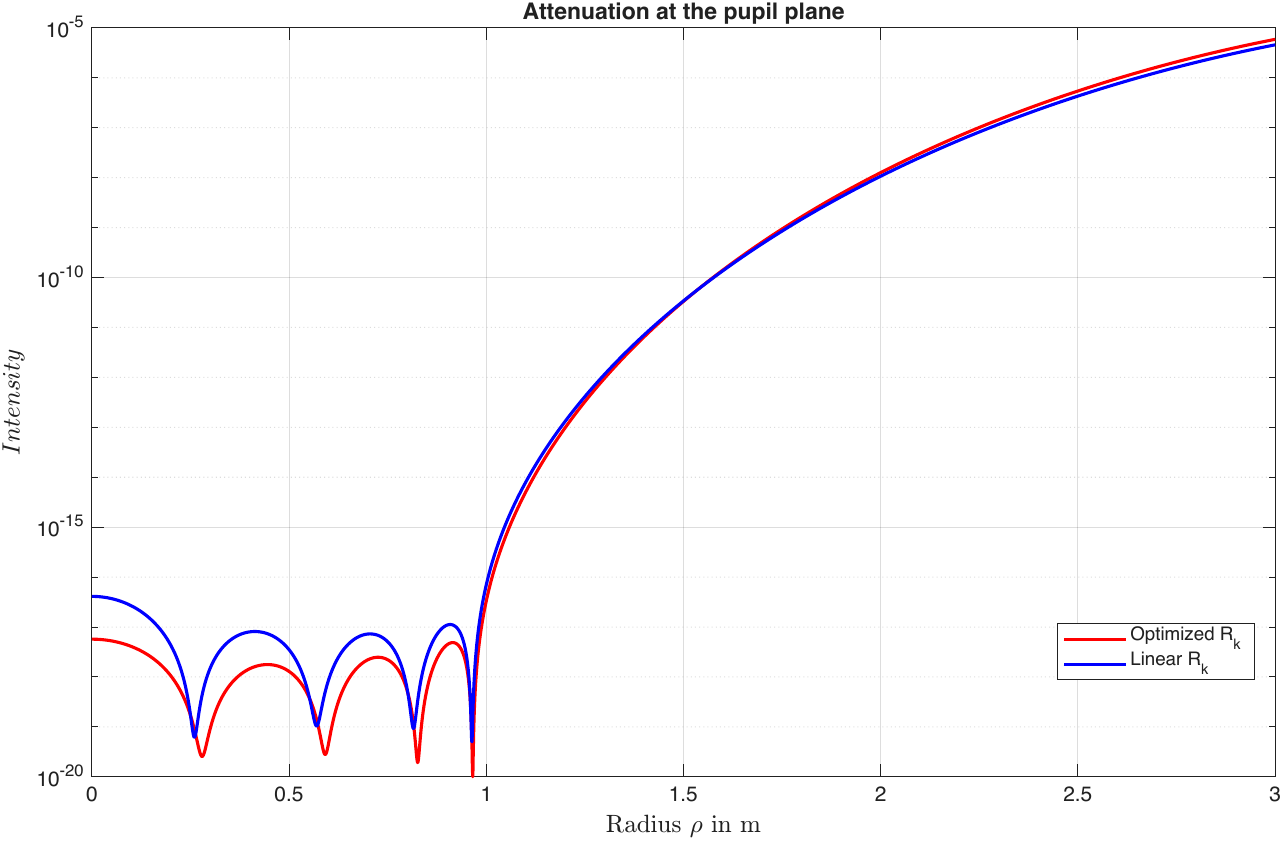}
 \\
    \includegraphics[width=0.84\linewidth]{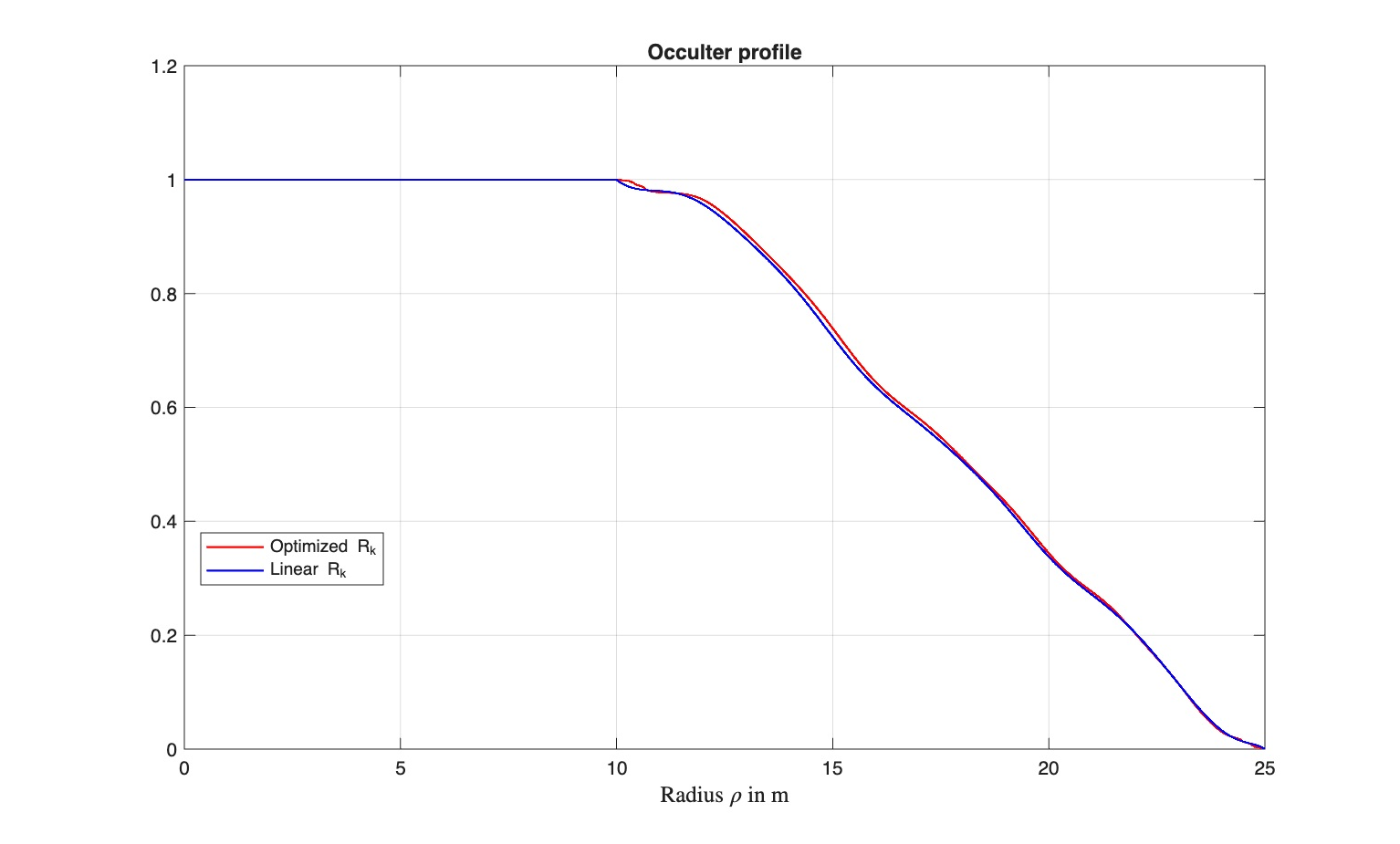}
    \caption{\emph{Top:} intensity obtained with linearly spaced   $R_k$ (blue) and with optimized $R_k$ (red), for $\mu_0 = 10^{-11}$. \emph{Bottom:} corresponding occulter profiles.}
    \label{Phas_Optim5}
\end{figure}

Numerical experiments show that only a few iterations on $\mathbf{R}$ and $\boldsymbol{\alpha}$ are sufficient. In this example, the attenuation is improved by about one order of magnitude when performing this alternating optimization. The resulting optimized intensity and corresponding occulter profiles are shown in Fig.~\ref{Phas_Optim5}.

   

\section{Petal shaped occulter}
\begin{figure}
     \centering
\includegraphics[width=0.7\linewidth]{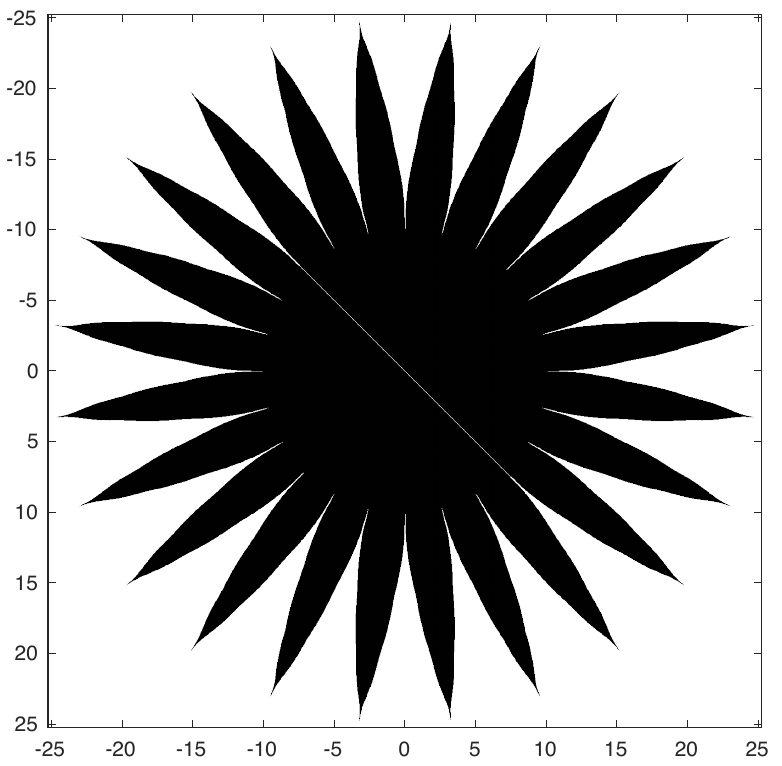}
\caption{Resulting petal-shaped .}
    \label{petals}
\end{figure}

In practice, an occulter with continuously varying transmission and a diameter of several tens of meters cannot be manufactured.
Petal-shaped occulters were proposed \cite{cash06} to reproduce, in azimuthal average, the transmission of apodized occulters. Fig.~\ref{petals}  shows the binary transmission $t(x,y)$ of the 24-petal occulter computed from the variable transmission $f(r)$ shown in the bottom part of Fig.~\ref{Phas_Optim5}.

For a sufficiently large number of petals \cite{vand07}, the central part of the diffraction pattern in the telescope pupil plane approaches that of the apodized occulter, where the telescope entrance pupil is located.

\section{Conclusion and outlook}
In this work, we showed that apodized occulter profiles can be efficiently optimized by expanding the transmission over a basis of circular occulters, for which the diffraction integral admits an analytic expression. The field computation relies here on the first term of the Lommel series, and extending the model to include additional terms is straightforward. 

Concerning the optimization, with fixed radii $R_k$, optimizing the coefficients $\boldsymbol{\alpha}$ amounts to a convex problem (quadratic objective with linear constraints), enabling reliable numerical solutions. Initial numerical experiments suggest that refining the radii $R_k$ via an alternating optimization between $\boldsymbol{\alpha}$ and $\mathbf{R}$ can provide additional attenuation; however, a more systematic study is needed to quantify this gain across configurations and bandwidths. 

Future work will focus on wide-band performance assessment and on the design of robust solutions accounting for mechanical tolerances.
\bibliography{biblio} 
\bibliographystyle{spiebib} 


\appendix 
\section*{Appendix: Gradient derivation}
\begin{enumerate}
\item For $m\neq k$ and $m\neq l$, $\partial\textbf{K}_{k,l}(\rho,\lambda)/\partial R_m=0$ 
\item $m=k$  

After some algebra:
\begin{multline}
    \frac{\partial\textbf{K}_{k,l}(\lambda)}{\partial R_m}=-\frac{2\pi}{\lambda z} \int_{0}^{\rho_M} \\ \left(R_m\sin(\pi\frac{R_m^2-R_l^2}{\lambda z})J_0(2 \pi \frac{\rho R_m}{\lambda z})+\rho \cos(\pi\frac{R_m^2-R_l^2}{\lambda z})J_1(2 \pi \frac{\rho R_m}{\lambda z})\right) \\ J_0(2 \pi \frac{\rho R_l}{\lambda z})2 \pi \rho d\rho
\end{multline}
\item $m=l$ 
\begin{multline}
    \frac{\partial\textbf{K}_{k,l}(\lambda)}{\partial R_m}=-\frac{2\pi}{\lambda z} \int_{0}^{\rho_M} \\ \left(R_m\sin(\pi\frac{R_m^2-R_k^2}{\lambda z})J_0(2 \pi \frac{\rho R_m}{\lambda z})+\rho \cos(\pi\frac{R_m^2-R_k^2}{\lambda z})J_1(2 \pi \frac{\rho R_m}{\lambda z})\right) \\ J_0(2 \pi \frac{\rho R_k}{\lambda z})2 \pi \rho d\rho
\end{multline}
equivalent to  $\frac{\partial\textbf{K}_{k,l}(\lambda)}{\partial R_m}$ since $\textbf{K}_{k,l}(\lambda)=\textbf{K}_{l,k}(\lambda)$ and using  $J'_0(x)=-J_1(x)$.
\end{enumerate}
Finally, with a factor 2 coming from the symmetrical terms $(k,l)$ and $(l,k)$:
\begin{multline}
\frac{\partial I}{\partial R_m}=-2 \frac{2\pi}{\lambda z} \sum_{k,l}\alpha_k \alpha_l  \int_{0}^{\rho_M} \\ \left(R_m\sin(\pi\frac{R_m^2-R_l^2}{\lambda z})J_0(2 \pi \frac{\rho R_m}{\lambda z})+\rho \cos(\pi\frac{R_m^2-R_l^2}{\lambda z})J_1(2 \pi \frac{\rho R_m}{\lambda z})\right) \\ J_0(2 \pi \frac{\rho R_l}{\lambda z})2 \pi \rho d\rho
\end{multline}

\end{document}